\documentclass{aip-cp}

\usepackage[numbers]{natbib}
\usepackage{rotating}
\usepackage{graphicx}

\newcommand{\PANDA}{$\overline{\textrm{P}}\textrm{ANDA}$}

\begin{document}

\title{New Spectroscopy with \PANDA~at FAIR: \\ X, Y, Z and the F-wave Charmonium States.}

\author[aff1] {Elisabetta Prencipe\corref{cor1}}
\author[aff2]{Jens S\"oren Lange}
\author[aff3]{Alexander Blinov}
\affil[aff1]{Forschngszentrum J\"ulich, J\"ulich (DE)}  
\affil[aff2]{Justus-Liebig-Universit\"at, Giessen (DE)} 
\affil[aff3]{Budker Institute of Nuclear Physics and Novosibirsk State University, Novosibirsk (RUS)}
\corresp[cor1]{e.prencipe@fz-juelich.de \\ on behalf of the \PANDA~collaboration.}

\maketitle

\begin{abstract}
Charm and charmonium physics have gained renewed interest in the past decade. Recent spectroscopic observations strongly motivate these studies. Among the several possible reactions, measurements in proton-antiproton annihilation play an important role, complementary to the studies performed at B-factories. The fixed target \PANDA~experiment at FAIR (Darmstadt, Germany) will investigate fundamental questions of hadron and nuclear physics in the interactions of antiprotons with nucleons and nuclei.  With reaction rates as large as 2$\times$10$^7$ interactions/s, and a mass resolution 20 times better as compared with the most recent B-factories, \PANDA~is in a privileged position to successfully perform the measurement of the width of narrow states, such as the $X(3872)$.  \PANDA~will investigate also high spin particles, whose observation was forbidden at B-factories, i.e. F-wave charmonium states.  In this report extrapolations on cross sections and rates with  \PANDA~are given.
\end{abstract}

\section{INTRODUCTION}

In the past decade many new, narrow states have been observed in the charmonium and bottomonium mass regions, which do not fit into a spectroscopical scheme as predicted by a static quark-antiquark potential model~\cite{models}. 

The $X(3872)$~\cite{x3872, x3872babar,  x3872cdf, x3872d0}, for example, is found very narrow and close to the $D D^*$ threshold. Recently the quantum numbers have been determined to $\rm{J^{PC}}=1^{++}$ by LHCb~\cite{x3872lhcb}. However, its nature is not understood, yet. The $Y(4260)$~\cite{y4260_babar, y4260_cleo-c, y4260_belle}  is found far above the open-charm thresholds; however no decay into $ \overline{D^{(*)}} D^{(*)}$ has been observed so far. Therefore it is being discussed e.g. as a possible hybrid with gluonic excitation. $Z$ states have raised attention after the discovery of the $Z(4430)$~\cite{lhcb4430, belle4430}, because  many of those $Z$ states are charged, which is in contradiction to conventional charmonium, inevitably being neutral. In the past 2 years several resonant structures, namely the $Z_c(3900)$~\cite{z3900_a, z3900_b}, the  $Z_c(4020)$~\cite{zc4020}, the $Z_c(3885)$~\cite{zc3885}, and the $Z_c(4025)$~\cite{zc4025}, have been observed. Their nature is  still unclear. 

The transition $Y(4260) \rightarrow Z(3900)^- \pi^+$ has been seen by BES III~\cite{ytoxbes}; the transition  $Y(4260) \rightarrow X(3872) \gamma$ has also been seen~\cite{ytoxgammabes}. But no experiment until now looked for the transition  $X \rightarrow Z$, or vice versa. Some Z states are observed decaying to $D D^*$ or $\overline{ D^*} D^*$. The mass values of the  $Z_c(3885)$, the $Z_c(3900)$, and the $Z_c(4020)$, published by BESIII,  are close to the $D D^*$ and $\overline{ D^*} D^*$ thresholds, respectively. Assuming that the Z states contain S-wave  $D D^*$ and $\overline{ D^*} D^*$ components, the spin parity  J$\rm ^P$of the  $Z_c(3885)$ and the $Z_c(3900)$ would be  J$\rm ^P$= $1^+$, and the spin parity of the $Z_c(4020)$ is expected to be J$\rm ^P$=$0^+$, $1^+$, or $2^+$. The former is confirmed by BESIII experimental data. One can expect also  similar S-wave resonances in the $\bar D D$ system, with J$\rm ^P$= $0^+$ (C=+1 for the neutral state), and mass values about 3730 MeV/c$^2$, which are not observed yet.

In this context, the contribution of a $\bar p p$ machine has to be considered as essential, because it can either confirm the above BES III measurements, and look for the non-observed $0^+$ Z states at the $\bar D D$ threshold, as  $\bar p p$ annihilation is a gluon rich process with direct access to various quantum numbers in production processes. In addition, the possibility of F-wave charmonium state search has been explored, as a test of flavor independence to understand the quark-antiquark potential.

\section{Estimates for the $X(3872)$ at \PANDA.}
The future \PANDA~experiment at FAIR (Facility for Antiproton and Ion Research) is well suited for charmonium studies, thanks to the high capability rate and the excellent mass resolution, that allows high precision measurements and energy scan. The experimental setup is described elsewhere~\cite{PB}. 

One of the most striking advantages of the \PANDA~experiment  is the opportunity to search for direct production of exotic resonant states with various quantum numbers, including  charged ones in $\bar p d$ collisions. In $e^+ e^-$ experiments only neutral $J^{PC} = 1^{--}$ resonances can be directly produced, and production of exotic charmed states through other mechanisms is suppressed. 

Using the detailed balance method, we can evaluate the cross section as:

\begin{equation}
\label{formula1}
\sigma[\bar p p  \rightarrow R ] \cdot BR( R \rightarrow f) 
= \frac{(2J + 1) \cdot 4\pi}{s- 4m^2_p}\cdot \frac{ BR ( R \rightarrow \bar p p) \cdot BR ( R \rightarrow f) \cdot \Gamma^2_{R}}{4(\sqrt s -m_{R})^2 + \Gamma^2_{R}}
\end{equation}

where $f$ is the final state of the decay channel,  $\Gamma$ is the total width of a resonance $R$, and $\sqrt s$ the center of mass energy.
For example, in order to evaluate the cross section of the process $\bar p p \rightarrow X(3872)$, we make use of the Equation~(\ref{formula1}), and obtain:

\begin{equation}
\label{formula1_x3872}
\sigma[\bar p p  \rightarrow X(3872)] \cdot BR(X(3872) \rightarrow f)
= \frac{3 \cdot 4\pi}{s- 4m^2_p}\cdot \frac{BR(X(3872) \rightarrow \bar p p) \cdot BR(X(3872) \rightarrow f) \cdot \Gamma^2_{X(3872)}}{4(\sqrt s -m_{X(3872)})^2 + \Gamma^2_{X(3872)}} . \\
\end{equation} 

We know that the spin parity of the $X(3872)$ is J$^{\rm P}$ = 1$^+$. We assume here a non-polarized incident beam. 
Down below we will use the decay channel $J$/$\psi$$\pi^+\pi^-$ as $f$ for the case of the $X(3872)$. 
If we run at $\sqrt s$ = $m_{X(3872)}$ = 3.872 GeV/c$^{\rm 2}$ the Equation~(\ref{formula1_x3872}) simplifies: $\sigma[\bar p p  \rightarrow X(3872)] = \frac{3\cdot 4\pi}{m^{2}_{X(3872)}- 4m^2_p}\cdot BR(X(3872) \rightarrow \bar p p)$. Here we assume c = $\hbar$ = 1.

The $BR((X(3872) \rightarrow \bar p p)$, then, enters the formula of Equation~(\ref{formula1_x3872}). We estimate it from the available experimental measurements in the PDG~\cite{pdg}, and those published by the LHCb experiment~\cite{detailbalanceLHCB}. The combination of both leads to an upper limit at 95$\%$ confidence level (c.l.):  $\sigma(\bar p p \rightarrow X(3872)) <$ (68 $\pm$ 0.4) nb. In agreement with theoretical predictions~\cite{braaten}, a reasonable number for the upper limit of the cross section $\sigma$($\bar p p \rightarrow X(3872)$) = 50 nb. Therefore, in \PANDA~we use to evaluate the expected $X(3872)$ yield  by using the above cross section estimate. This value should be interpreted as an upper limit. A lower limit estimate to the $X(3872)$ cross section cannot be quoted yet, simply because its very narrow width leads to unreasonable lower limits, by using standard methods for cross section evaluations.

\PANDA~could start in different operation modes, involving different antiproton beam resolution and luminosity values. Assuming the $X(3872)$ cross section in $\bar p p$ annihilation equal to 50 nb~\cite{braaten, soerencharm2013}, we  are expected to produce 432000 $X(3872)$ per day in high luminosity mode (average luminosity $\cal L $ = 10$^{\rm 32}$ cm$^{\rm -2}$ s$^{\rm -1}$), and 43200 $X(3872)$ per day in high resolution mode (average luminosity $\cal L$ = 10$^{\rm 31}$ cm$^{\rm -2}$ s$^{\rm -1}$).  Thus, \PANDA~can be considered as a ''mini-$X(3872)$ factory''. In the latter situation, the mass scan in 100-keV-steps, that is needed to measure the $X(3872)$ width,  can be performed in about 3 weeks, collecting 15 points above and below the mass threshold, as detailed explained in Ref.~\cite{martin}. In high resolution mode, \PANDA~is designed to have $\Delta p /p$ = 5 $\cdot 10^{-5}$.  

\section{Estimates for the $Y(4260)$ at \PANDA.}

\label{estimateY}

We calculate the number of produced $Y(4260)$ by multiplying the expected luminosity and the cross section of the process $\bar p p \rightarrow Y(4260)$. 
We assume $BR(Y(4260) \rightarrow J/\psi \pi^+ \pi^-)$ = 100$\%$, for four reasons:

\begin{itemize}

\item the decay $Y(4260) \rightarrow J/\psi \pi^+ \pi^-$ was the discovery mode \cite{y4260_babar};

\item for all known $Y(4260)$ decay channels, the PDG~\cite{pdg} quotes ``seen''
with no numbers reported; 

\item 
all searches for decays to open charm performed at B factories, in ISR and B decay modes, 
lead to upper limits only. In the PDG~\cite{pdg}, these upper limits are all normalized 
to the BR($Y(4260) \rightarrow J/\psi \pi^+ \pi^-$)~\cite{y1, y2, y3, y4, y5, y6, y7, y8};

\item recently, the BESIII experiment published the observation of the transition $Y(4260) \rightarrow \gamma X(3872)$~\cite{ytoxgammabes}, from which it can be concluded that the $BR(Y(4260) \rightarrow \gamma X(3872)$, with $ X(3872) \rightarrow J/\psi \pi^+ \pi^-$, contributes in negligible way to the total $BR(Y(4260))$, i.e.\  $\leq$0.5$\%$ only. 

\end{itemize}

The cross section of the process $\bar p p \rightarrow Y(4260)$ can be estimated using detailed balance (Equation~(\ref{formula1})). 
However, if the poorly known upper limit 
$BR(Y(4260) \rightarrow \bar p p) / BR(Y(4260) \rightarrow J/\psi \pi^+ \pi^-) <$0.13 at 90$\%$ c.l.~\cite{babaruly}
is taken as estimate and inserted into Equation~(\ref{formula1}),  
it leads to an unrealistically high cross section estimate of 4370 nb. 

If the BR of a resonant state, decaying to $\bar p p$, is known, we can directly apply the detailed balance method to evaluate the cross section. However, if this BR is not known, at first we use the ansatz that partial width is identical for states $R_1$, $R_2$ of the same quantum number~\cite{PB}. 

\begin{equation}
BR ( R_1 \rightarrow p \overline{p} ) =  
BR ( R_2 \rightarrow p \overline{p} ) \cdot \frac{\Gamma_{total} ( R_2 ) }{\Gamma_{total} ( R_1 )}.
\label{escaling}
\end{equation}

This method assumes that the partial widths $\Gamma (R \rightarrow \bar p p)$ of all charmonium states are identical, 
where $R$ refers to the state. Although we might have indication that the $Y(4260)$ is not a charmonium state, no model exists to evaluate the cross section for exotic states. In absence of any explanation of the $Y(4260)$ nature, thus we perform our calculation under the naive assumption that it is a charmonium state.
As a reference state for the $Y(4260)$ estimates, we choose the $\psi(3770)$, for which the following numbers have been recently measured~\cite{bes3cross}: $BR(\psi(3770) \rightarrow \bar p p) = (7.1^{+8.6}_{-2.9}) \cdot 10^{-6}$, and $\sigma(\bar p p \rightarrow \psi(3770)) = (9.8^{+11.8}_{-3.9} )~ nb$. 
We start our calculation from Equation~(\ref{escaling}), using $\psi$(3770) as a reference, and obtain: 

\begin{equation}
\label{myscaling}
BR ( Y(4260) \rightarrow p \overline{p} ) = BR ( \psi(3770) \rightarrow p \overline{p} ) \cdot \frac{\Gamma_{total} ( \psi(3770) ) }{\Gamma_{total} ( Y(4260) )}.
\end{equation}
We can write again Equation~(\ref{myscaling}), using the detailed balance principle, as: 

\begin{equation}
\label{my2}
\sigma ( p \overline{p} \rightarrow Y(4260) ) = \sigma ( p \overline{p} \rightarrow \psi(3770) ) \cdot \frac{\Gamma_{total} ( \psi(3770) ) }{\Gamma_{total} ( Y(4260) )} = 9.8~nb  \cdot \frac{27.2~MeV}{102~MeV} =  2.2~nb.
\end{equation}
We assume the cross section of Equation~(\ref{my2})  as an upper limit.
In order to estimate a lower limit for the cross section $\bar p p$$\rightarrow$$Y(4260)$, we use the assumption that the annihilation part, which manifests in the decay into $e^+e^-$ and the decay into $\bar p p$, are identical:

\begin{equation}
\sigma ( p \overline{p} \rightarrow Y(4260) ) = 2.2~nb \cdot \frac{\Gamma_{ee} ( Y(4260) )}{\Gamma_{ee} ( \psi(3770) )} =
2.2~nb \cdot \frac{\Gamma_{ee} ( Y(4260) )}{BR( \psi(3770)\rightarrow e^+ e^- ) \cdot \Gamma_{total} ( \psi(3770) )}  = 0.077~nb.
\label{escaling_ee_pp}
\end{equation}

This result is obtained by using the partial width $\Gamma_{ee}$($Y(4260)$) and $\Gamma_{total}$ from PDG~\cite{pdg},  and $BR$($\psi(3770) \rightarrow \bar p p $) from Ref.~\cite{bes3cross}. As a word of caution, the scaling in Equation~(\ref{myscaling}) is only an approximation as well, as $e^+$ and $e^-$ are point-like particles,
but $p$ and $\bar p$ are not. When scaling the partial width $\Gamma (R \rightarrow \bar p p)$ 
(or the branching fraction $BR (R \rightarrow \bar p p)$) for a decay to $\bar p p$ of a resonance $R$=$R_1$ 
with a mass $m_1$ to another resonance $R$=$R_2$ with a mass $m_2$, one would have to take into account, that the proton formfactor $G$ has an energy dependence $G$($\sqrt{s}$)
and is changing from $\sqrt{s}$=$m_1$ to $\sqrt{s}$=$m_2$. However, we do not have to apply this correction here for the evaluation of the lower cross section limit for the $Y(4260)$, as the formfactor is already implicitely included in the measured $BR ( \psi(3770) \rightarrow \bar p p)$). 

The cross section $\sigma(\bar p p \rightarrow Y(4260))$ could be compared to the cross section $\sigma(e^+e^- \rightarrow  Y(4260))$=(62.9 $\pm$ 1.9) pb~\cite{z3900_b}. For the case of the above upper limit, the cross section in the $\bar p p$ process is a factor of about 35 larger than the cross section measured in  $e^+e^-$ collisions.

\subsection{Estimation of the produced $Z_c(3900)$ at \PANDA.}

The number of expected $Z_c(3900)$ events in \PANDA~can be estimated from
Refs.~\cite{ytoxbes}, \cite{soerencharm2013}, and \cite{elisabetta}:
in the decay $e^+e^- \rightarrow J/\psi \pi^+ \pi^-$ the BESIII experiment
observed the $Z_c(3900)$~\cite{ytoxbes},
using the full dataset collected near the $Y(4260)$ energy. 
The observed $Z_c(3900)$ yield is 307, and the ratio
\begin{equation}
R = \frac{\sigma(e^+e^- \rightarrow Z_c(3900)^+ \pi^- \rightarrow J/\psi
\pi^+ \pi^-)}{\sigma(e^+e^- \rightarrow J/\psi \pi^+ \pi^-)} = 21.5\% \ .
\end{equation}

All measurements are based on 1.9 fb$^{-1}$, which  is presently the world
largest dataset collected at the $Y(4260)$ energy. We can extrapolate how many produced $Z_c(3900)$ are expected at \PANDA, assuming that [66$-$1900] $Y(4260)$ states are expected to be produced per day in high resolution mode. The calculation is based upon the cross section range [77$-$2200] pb, as stated in the previous Section. 

For the upper limit evaluation we find:

\begin{equation}
\sigma(\bar p p \rightarrow Z_c(3900)) = \sigma(\bar p p \rightarrow
Y(4260)) \cdot 21.5 \% = 0.473 ~nb \ .
\end{equation}

For the lower limit evaluation we find:

\begin{equation}
\sigma(\bar p p \rightarrow Z_c(3900)) = \sigma(\bar p p \rightarrow
Y(4260)) \cdot 21.5 \% = 0.017 ~nb \ .
\end{equation}

Based on these estimates, we would be able to produce [14$-$405] $Z_c(3900)$
events/day in high resolution mode,
when running at a center-of-mass energy for $Y(4260)$ peak resonance
production. We note that of course there can be non-resonant production $\bar p p \rightarrow J/\psi \pi^+ \pi^-$
at the same energy, and the according non-resonant cross section can be
even larger than the resonant cross section,
although the indication from BESIII and the ISR measurements at the B
factories is that non-resonant $e^+ e^- \rightarrow J/\psi \pi^+ \pi^-$
is small (${\cal O}$$<$10\%).

\section{$Z \leftrightarrow X$ transitions at \PANDA.}

Observations of transitions of X, Y and Z states are very important for
understanding the spectroscopical pattern,
and possibly conclude similarities in the nature of these states.
Two recent BESIII publications connect the $X(3872)$ to the
$Y(4260)$~\cite{ytoxgammabes},
and the $Y(4260)$ to the $Z_c(3900)$~\cite{ytoxbes}.
However, up to now, no experimental measurement connects the $X(3872)$ to
the Z structures.
Thus, we propose to search for the transitions X to Z (or Z to X).
\PANDA~would be well suited for this search because of the following
reasons:
\begin{itemize}

\item about possible Z to X transitions, the decay
$Z(3900) \rightarrow X(3872) \pi$ is kinematically forbidden.
The decay $Z(4020) \rightarrow X(3872) \pi$ is allowed, however
suppressed as a P-wave decay close to threshold, since
both the $X(3872)$ and the $Z(4020)$ have positive parity (assuming S-wave $\overline{D^*} D^*$ content of the $Z(4020)$). Two pion transitions between the $Z(4020)$ and the $X(3872)$ would go in S-wave, but they are kinematically forbidden;

\item $Z(4020)^+ \rightarrow J/\psi \pi^0 \pi^+$ is allowed, but no signal
for the $Z(4020)$ was observed in the investigation of the according final state in searches for the charged partner of the $X(3872)$ at BaBar~\cite{nochargedXBaBar} and Belle~\cite{nochargedXbelle};

\item \PANDA~will collect a data set of $X(3872)$ (see above), with a
statistics larger than other experiments by one or two orders of magnitude. Thus, rare decays of the X(3872), e.g. isospin forbidden decays or radiative decays,
become accessible;

\item as the $Z(3900)$ was observed in close vicinity of the $D$$D^*$
threshold, and the $Z(4020)$ was observed in close vicinity
of the $D^*$$D^+$ threshold, it is intriguing to assume the existance of
another yet unboserved $Z(3730)$ in close vicinity
of the $\bar D$$D$ threshold. Assuming S-wave, this state would have
$J^P$=$0^+$, and thus it cannnot decay to $J$/$\psi$$\pi$
due to parity conservation. In fact, neither charged or neutral structure
have been observed around this mass in this final
state. Using the future $X(3872)$ data sample at \PANDA, $X(3872) \rightarrow Z(3730) \pi$ represents a candidate decay channel. The latter decay is suppressed due to isospin violation; however, isospin violating decays of the $X(3872)$, such as $X(3872)$$\rightarrow$$J$/$\psi$$\rho$, have been observed with significant branching fractions. In addition, the requirement of a $J$/$\psi$ in the final state provides a tool to reduce the hadronic background at \PANDA. Simulations performed at the $X(3872)$ energy scan have already shown that the ratio signal over background is 6:1~\cite{martin}; therefore, a favorable ratio S/B is expected also for the search of the $Z(3730)$ resonant structures;
\item due to the observation of the $Z(3900)^0$ and the $Z(4020)^0$,
$Z_c$ states have been interpreted as isospin triplets with charged and
neutral partners at the same mass. Thus, we may search for the $Z(3730)^0$, which could be reconstructed from $J/\psi \gamma$ and $\chi_{c1} \pi^0$ decays.
In fact, in these transitions the parity flips from J$^{ \rm P}$ = 1$^+$ (the
$X(3872)$) to J$^ {\rm P}$ = 0$^+$$\times$0$^-$. Although radiative decays are suppressed by $\alpha$/$\pi$, the observation of this decay would be of very high importance, as it would provide a way to measure the C-parity of the $Z(3730)^0$;

\item in an additional stage, we could also search for the charged
$Z(3730)^+$ candidate, decaying to $\chi_{c1} \pi^+$, with subsequent $\chi_{c1} \rightarrow J/\psi \gamma$ and $J/\psi \rightarrow$ leptons (with L=1).
Investigation of other final states, e.g. $Z(3730) \rightarrow$$D$$\overline{D}$
in e.g.\ $p$$\overline{p}$$\rightarrow$$D$$\overline{D}$$\pi$ are also
possible, but would suffer of higher background. Again, it should be noted that a dedicated data taking run at the center of mass energy of $Z(3730)$ is not required for the proposed study.

\end{itemize}

To summarize, \PANDA~would be unique to search for $X \rightarrow Z \pi$
transitions involving
yet unobserved neutral and charged $Z(3730)$ states in the processes:

\begin{itemize}
\item $\bar p p \rightarrow Z(3730)^0 \pi^0$ , $Z(3730)^0 \rightarrow
J/\psi \gamma$, with $J/\psi \rightarrow$ leptons and $\pi^0 \rightarrow
\gamma \gamma$;\\
\item $\bar p p \rightarrow Z(3730)^0 \pi^0$ , $Z(3730)^0 \rightarrow
\chi_{c1} \pi^0$, with $\pi^0 \rightarrow \gamma \gamma$, $\chi_{c1}
\rightarrow J/\psi \gamma$ and $J/\psi \rightarrow$ leptons;\\
\item $\bar p p \rightarrow Z(3730)^\pm \pi^\mp$ , $Z(3730)^\pm
\rightarrow \chi_{c1} \pi^\mp$, with  $\chi_{c1} \rightarrow J/\psi
\gamma$  and $J/\psi \rightarrow$ leptons.
\end{itemize}

We also note, that the $Z_c(3900)$ and the $Z_c(4020)$ have not been observed in
$B$ decays, yet. Thus, we expect high discovery potential for \PANDA.

\section{F-wave charmonium states}
A unique feature for \PANDA~can be the search for high spin states. Based on theoretical predictions as in Ref.~\cite{swanson}, we simulated the multiple radiative cascade $1 ^3F_4 (J^{PC} = 4^{++}) \rightarrow 1 ^3D_3 (J^{PC} = 3^{--})\rightarrow \chi_{c2} (J^{PC} = 2^{++})\rightarrow J/\psi (J^{PC} = 1^{--})$, as detailed reported in Ref.~\cite{soerencharm2013}. \PANDA~is designed to perform an excellent photon reconstruction, and our simulations have already demonstrated that physics channels reconstructed from one $J/\psi$ and three $\gamma$ have clear signature, and a background suppression factor of about 10$^6$~\cite{soerencharm2013}. The static heavy quark anti-quark ($\bar Q Q$) potential of the Cornell-type ~\cite{swanson, cornell0, cornell1} can be expressed by $V(r) = \frac{4}{3} \frac{\alpha_s}{r} + k \cdot r$,  with a chromo-electric Coulomb-type term, and a linear confinement term. It predicts many of the experimentally observed charmonium and bottomonium states up to a  precision of $\approx$1 MeV. Recently several new states have been observed, which fit well into the prediction of the Cornell-type potential, i.e. the $h_c$, the $h_b$ and the $h_b '$, or $\eta_b$ and $\eta_b '$.  By the mass measurements of these new states, a comparison of the level spacings between charmonium (mass region 3-4 GeV/$c^2$) and bottomonium (mass region 9-10 GeV/$c^2$) became available for the first time. For example, the following spin-averaged mass differences:

\begin{equation}
m(h_c) - \frac{3 \cdot m(J/\psi) + m(\eta_c)}{4} =
456.8 \pm 0.2 ~MeV
\end{equation}

\begin{equation}
m(h_b) - \frac{3 \cdot m(\Upsilon(1S) + m(\eta_b)}{4} =
453.5 \pm 1.4 ~MeV
\end{equation}
are identical to a level of better than 10$^{-3}$, which is quite
surprising and points to flavor independence of the quark anti-quark
potential. In other words, the potential does not seem to depend on the
different quark mass of the charm or the bottom quark, although in the
Cornell potential the quark mass is explicitly one of the adjustable
parameters.

\begin{table*}
\caption{Summary of the expected X, Y, and Z production rates per day in \PANDA, assuming different operation modes (e.g. different rates $\cal {L}$/day). The calculation is performed by multiplying luminosity and cross sections. The  cross section upper limits are used in these calculations, and in parenthesis the corresponding lower limit is reported. For the $X(3872)$, only an upper limit was evaluated in this short report, and thus we omit a second number.}
\label{table2}
\begin{center}
\vskip -0.2cm
\begin{tabular}{lrccccl}
\hline  
  Resonance      & $\cal L$ = 8.64 $pb^{-1}/day$& $\cal L$= 0.864 $pb^{-1}/day$& $\cal L$= 0.432 $pb^{-1}/day$ \cr \hline
  $X(3872)$    & 432000 & 43200& 21600 \cr
$Y(4260)$    & 19000 (665)&1900  (67)& 950 (7) \cr
$Z(3900)^+$    & 4050 (140)& ~405 (14)&202 (7) \cr

 \hline
\end{tabular}
\end{center}
\end{table*}
However, as already found in the 1970's~\cite{cornell2}, flavor independence is not fulfilled for a Cornell-type potential. Potentials, for which identical level spacings for charmonium and bottomonium are fulfilled, are logarithmic potentials of the type $V(r) = r_1 ln (c_2 r)$.

One of the important tasks of future experiments such as \PANDA~is the search for additional, yet unobserved states (e.g. the $h_c '$ or $^3 F_4$ state), which could be used to obtain additional level spacings and further test the
flavor indepedence, and possibly a logarithmic shape of the potential. Simulations in this sense were performed with the \PANDA~full reconstruction framework~\cite{pandaroot}, and they are promising, as detailed in Ref.~\cite{soerencharm2013, elisabetta}.
\section{Summary}

In summary, Table~\ref{table2} reports our estimates for X, Y, Z production rates at \PANDA, assuming different luminosity average values $L$ = 10$^{\rm 32}$ cm$^{-2}$s$^{-1}$, $L$ = 10$^{\rm 31}$ cm$^{-2}$s$^{-1}$, and $L$ = 0.5$\times$10$^{\rm 31}$ cm$^{-2}$s$^{-1}$, respectively. Rates must be interpreted as educated guess, due to the $\bar p p$ cross sections, which have not been measured, yet. The expected large statistics at \PANDA~will help to address many open questions about X, Y, and Z states, in order to unravel their nature. A high discovery potential exists, in particular for new states with quantum numbers unobservable in production processes at other experimental facilities. 

\bibliographystyle{aipnum-cp}%

\end{document}